\documentclass[preprint,prb,showpacs,amsmath,amssymb, floatfix]{revtex4}
\usepackage{bm}
\usepackage{graphicx}

\def\Idc{{I_{\textrm{dc}}}}
\def\rxx{{ r_{xx}  }}

\begin{document}

\title{Non-linear magnetotransport phenomena in high-mobility two-dimensional electrons in InGaAs/InP and GaAs/AlGaAs }
\author{S. A.~Studenikin}
\email{sergei.studenikin@nrc.ca}
 \affiliation{National Research Council of Canada, Ottawa, Ontario K1A 0R6, Canada}
\author{G. Granger}
  \affiliation{National Research Council of Canada, Ottawa, Ontario K1A 0R6, Canada}
\author{A. Kam}
 \affiliation{National Research Council of Canada, Ottawa, Ontario K1A 0R6, Canada}
\author{A. S. Sachrajda}
 \affiliation{National Research Council of Canada, Ottawa, Ontario K1A 0R6, Canada}
\author{Z. R.~Wasilewski}
 \affiliation{National Research Council of Canada, Ottawa, Ontario K1A 0R6, Canada}
\author{P. J. Poole}
 \affiliation{National Research Council of Canada, Ottawa, Ontario K1A 0R6, Canada}

\date{\today}
\begin{abstract}
  This paper reports on the observation and analysis of magnetotransport phenomena in the nonlinear differential resistance $r_{xx}=dV_{xx}/dI$ of high-mobility In$_x$Ga$_{1-x}$As/InP and GaAs/Al$_x$Ga$_{1-x}$As Hall bar samples driven by direct current, $\Idc$.
  Specifically, it is observed that Shubnikov -de Haas (SdH) oscillations at large filling factors invert their phase at sufficiently large values of $\Idc$.
  This phase inversion is explained as being due to an electron heating effect.   
  In the quantum Hall effect regime the $r_{xx}$ oscillations transform into diamond-shaped patterns with different slopes corresponding to odd and even filling factors. The diamond-shaped features at odd filling factors can be used as a probe to determine spin energy gaps.
  A Zero Current Anomaly (ZCA) which manifests itself as a narrow dip  in the $r_{xx}(\Idc)$ characteristics at zero current, is also observed.  The ZCA effect strongly depends upon temperature, vanishing above 1 K while the transport diamonds persist to higher temperatures. 
  The transport diamonds and ZCA are fully reproduced in a higher mobility GaAs/AlGaAs Hall bar structure confirming that these phenomena reflect intrinsic properties of two-dimensional systems.

\end{abstract}
 \maketitle

\section{Introduction.}
\noindent Nonlinear transport phenomena in high mobility two-dimensional electron gas (2DEG) systems have recently received considerable attention, motivated originally by the observation of microwave-induced resistance oscillations which under certain conditions can evolve into zero resistance states in high mobility GaAs/AlGaAs samples.\cite{zudov01,ye,mani02,zudov03}
These studies have led to other interesting experimental observations, such as  nonlinear effects driven by direct current (dc), for example, Hall-effect induced resistance oscillations, zero differential resistance states, and current instabilities.\cite{Du02,zudov07,vitkalov1}
Theoretically, it has been shown that the observed microwave and dc bias driven nonlinear phenomena in high mobility 2DEG samples originate from the non-equilibrium distribution function and Landau quantization.\cite{dmitriev1,fedorych,dmitrievCondMat}
Currently, the above mentioned nonlinear phenomena have been studied exclusively in extremely high mobility GaAs/AlGaAs structures. On the other hand, driven partly by quantum information proposals, there is an increasing interest in InGaAs structures which offer a larger electron g-factor and strong spin-orbit coupling.\cite{schapers1,samuelson1,pepper1,daniele,majorana}

In this work we discuss yet other non-linear phenomena: strong distortion of Shubnikov-de Haas (SdH) oscillations leading to phase inversion, and  Zero Current Anomaly (ZCA).
Note, that  the SdH phase inversion  has not been reported in earlier magneto-transport experiments exploring electron heating effects in quantizing magnetic fields.\cite{leadley,fletcher}
Only recently, two publications have appeared on the observation of SdH phase inversion under  dc bias in a GaAs/AlGaAs structure \cite{Kalmanovitz} and in a graphene sample.\cite{Tan11}
In a high mobility GaAs/AlGaAs structure \cite{Kalmanovitz}  (mobility $\simeq$850,000 cm$^2$/Vs) the phase inversion is discussed in connection with a zero differential resistance effect due to a strong non-linearity attributed to a spectral diffusion phenomenon.
In a low mobility graphene sample \cite{Tan11} (electron mobility of about 8,000 cm$^2$/Vs) a SdH phase inversion was quantitatively described by the Isihara formula\cite{Isihara}  taking into account an electron gas heating effect.

The focus of this paper is to investigate nonlinear effects in the differential resistance of high mobility In$_x$Ga$_{1-x}$As/InP Quantum Well (QW) structures over a wide magnetic field range. In particular, we extend our study into the Quantum Hall Effect (QHE) regime which has not been accessible previously, in ref.~[\onlinecite{Kalmanovitz}] due to instabilities that were caused by the too high electron mobility and concentration, while in the graphene sample\cite{Tan11} the mobility was too low.
Two novel phenomena are observed:
(i) a phase inversion of the SdH oscillations  which at higher magnetic fields transform into diamond-shaped features and can be linked to previous QHE breakdown experiments;
(ii) a Zero Current Anomaly (ZCA) which manifests itself as a narrow dip in the differential resistance at zero dc bias.
To confirm the general nature of the observed phenomena, we repeat these measurements on a standard high-mobility GaAs/Al$_x$Ga$_{1-x}$As heterostructure and indeed find the same effects in the both material systems.
\begin{figure}
\includegraphics[width=0.8\columnwidth]{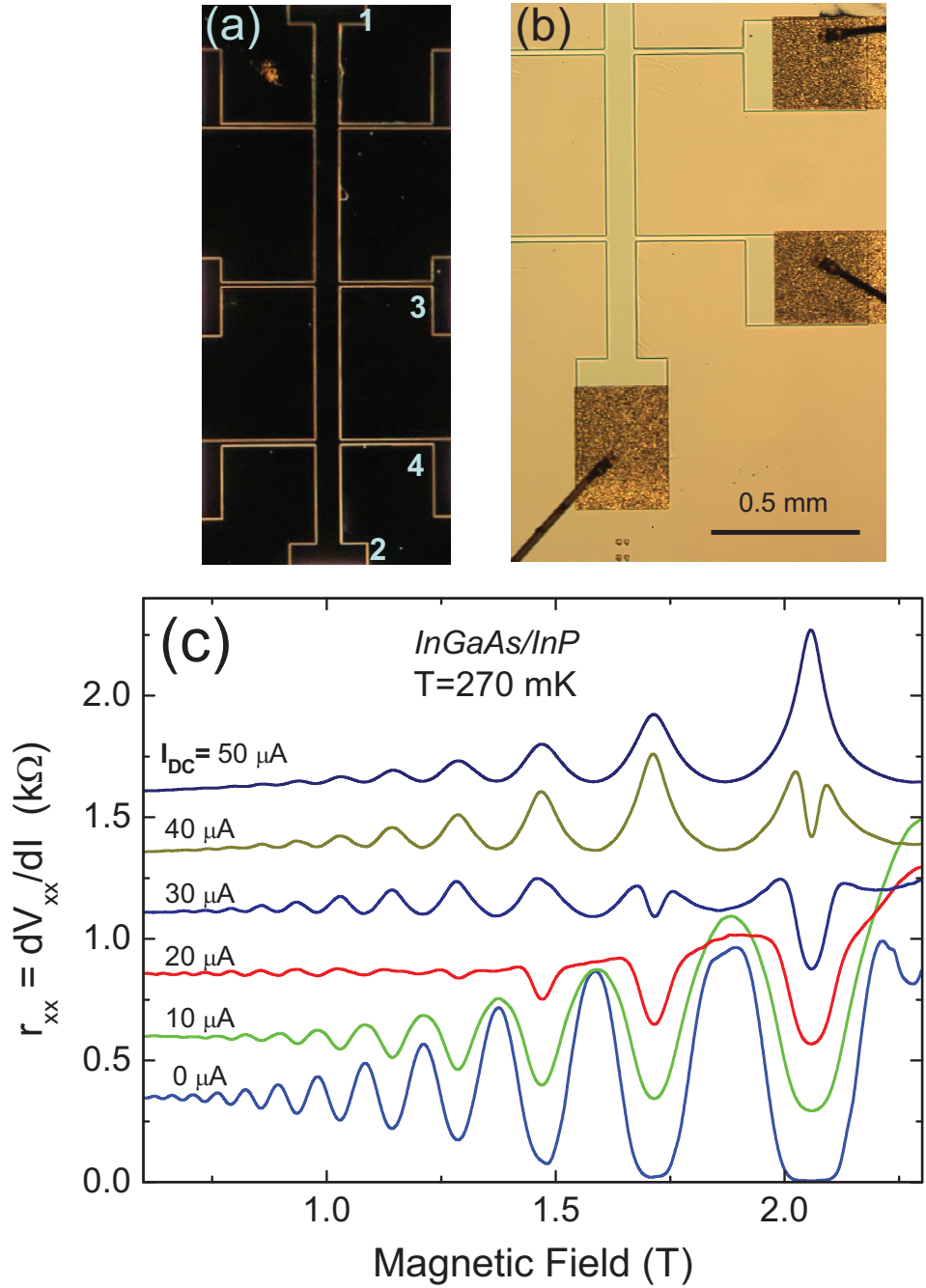}
\caption{ (a,b) Optical images of the w=100 $\mu m$ width Hall bar sample at different magnifications taken by microscopes (a) Olympus SZ61, and (b) Olympus BH-2.
(c) Differential magnetoresistance traces of a 100 $\mu m$ wide Hall bar sample fabricated from an InGaAs/InP QW structure at different DC values, T=270 mK. All curves except at $\Idc=0$ are shifted vertically by 0.25 k$\Omega$ progressively for clarity.}
 \label{Fig1}
\end{figure}

\section{Experimental Details.}
 The InGaAs/InP QW  high-mobility structure is grown on a semi-insulating (001) InP substrate by Chemical Beam Epitaxy (CBE) using  Trimethyl-Indium In(CH$_3$)$_3$ and Triethyl-Gallium Ga(C$_2$H$_5$)$_3$ as the In and Ga precursors correspondingly, and  AsH$_3$ and PH$_3$ as the As and P precursors.\cite{philip,ghislain}
 After loading in the CBE growth chamber, the substrate surface is first thermally cleaned at $580^{\circ}$C.
 The following layers are then grown at a substrate temperature of $515^{\circ}$C: a 17 nm undoped InP buffer layer, a 10 nm undoped In$_x$Ga$_{1-x}$As (x=0.76) layer forming the QW conduction channel, followed by a 20 nm InP undoped spacer, a 5 nm Si-doped InP layer at the volume density of $6.7\times 10^{17}$ cm$^{-3}$ and, finally, a 44 nm undoped InP cap layer.
For the nonlinear transport measurements a 100~$\mu$m wide Hall bar is prepared by means of standard optical lithography and wet etching.
The Ohmic contacts are based on a NiAuGe eutectic using standard rapid annealing in an Ar atmosphere at 420$^{\circ}$C for 5 minutes.
Two optical images of the Hall bar sample are shown in Fig. 1 at different magnifications taken with different microscopes.
The Hall bar and potential arms in Fig. 1 (b) have widths of  w=100~$\mu$m and  10 $\mu$m, respectively.   The narrow potential arms are also used for test measurements, e.g. to exclude contact effects and to verify how the observed phenomena depend on the Hall bar width.

 The Hall bar is mounted on a sorption pumped $^3$He Oxford Instruments Heliox insert equipped with a 5T superconducting magnet.
 Four-point  measurements of the differential resistance, $r_{xx}(\Idc)=dV_{xx}/dI$, are performed by applying a combination of a dc and a small ac ($I_{\textrm{ac}}\leq$ 50 nA) modulation through the current contacts (1 and 2 in Fig. 1 (a)) and  measuring the ac voltage drop, $dV_{xx}$ between the potential contacts (3 and 4 in Fig. 1 (a)) using  standard lock-in technique.

After illumination with a red light emitting diode, the 2DEG  attains  a concentration of 5.3$\times$10$^{11}$ cm$^{-2}$ (corresponding to a Fermi energy of $\varepsilon_F$=28 meV) and a mobility of 190,000~cm$^2$/Vs, respectively, measured at 2~K.\footnote{Currently, in our highest quality InGaAs/InP QW samples the electron mobility reaches $\mu=3.1\times 10^{5}$ cm$^2$/Vs at a concentration of 4.3$\times$10$^{11}$ cm$^{-2}$.}
Several traces of $r_{xx}$ measured on the 100 $\mu m$ wide InGaAs/InP Hall bar at different values of $\Idc$ are plotted in Fig.~1 (c).  Regular SdH oscillations are observed at $\Idc$=0 (bottom trace), whereas with increasing dc magnitude the SdH minima progressively evolve into maxima and by $\Idc$$\simeq$50~$\mu A$  all the SdH minima have inverted into peaks.
Let us start with examining the low field range (large filling factors) where the SdH amplitude $\Delta R_{xx}$ is small compared to the zero-field resistance $R_0$.

\section{Experiment and simulation in small magnetic fields.}
In the small magnetic fields regime the SdH oscillations in resistance $R_{xx}=V_{xx}/I$  can be described analytically by the following equation:\cite{Isihara,Coleridge}
\begin{equation}
\Delta R_{xx} = 4 R_0 D_T(X_T) \exp(-D_q/\hbar \omega_c)\cos (\frac{2\pi  E_F}{\hbar \omega_c}-\pi),
\label{eq.1}
\end{equation}
where only the first harmonic of the Fourier expansion is retained, $R_0$ is the zero field resistivity, $D_{q}=\pi \hbar/\tau_q$ is the Dingle damping parameter,  $\tau_q$ is the quantum lifetime, $E_F$ is electron Fermi energy, $D_T = X_T /\sinh(X_T)$ is the thermal damping factor with $X_T = 2\pi^{2} k_BT / \hbar\omega_{c}$, $k_B$ being the Boltzman constant,   and $\hbar\omega_{c}$ the cyclotron energy.
The above equation is derived using Lorentzian shape of Landau Levels (LL) to calculate electron Density of States (DOS).
It should be noted that different LL line shapes have been considered in the literature, in particular, Lorentzian, Gaussian, and SCBA (self-consistent Born approximation) types.\cite{Isihara,Coleridge,Endo,Dmitriev2}
For Lorentzian levels the normalized shape of one LL is given by  $\nu_1(E)=1/\pi\Gamma_L/[1+(E/\Gamma_L)^2]$
with a full width at a half maximum FWHM = 2$\Gamma_L$, where $\Gamma_L=\hbar/2\tau_q$ with B-independent $\tau_q$.
Summation over all Lorentzian LLs results in the Isihara formula  valid for an arbitrary DOS modulation.\cite{Isihara}
In the other two cases of a Gaussian\cite{Endo}, $\nu_1(E)=1/(\sqrt{2\pi}\Gamma) \exp(-E^2/2\Gamma^2)$,  and a SCBA  semi-elliptical Landau levels\cite{Dmitriev2}, the width of LLs is proportional to $\sqrt{B}$, namely $\Gamma_G=\sqrt{\hbar \omega_c \Gamma_L/\pi}$.  Our numerical simulations of the DOS show that all the three above-mentioned cases coincide exactly with each other at low magnetic fields where eq.(1) is valid.  Therefore, we will use eq. (1) in our analysis and modeling the  SdH phase inversion effect.

\begin{figure}
\includegraphics[width=1.0\columnwidth] {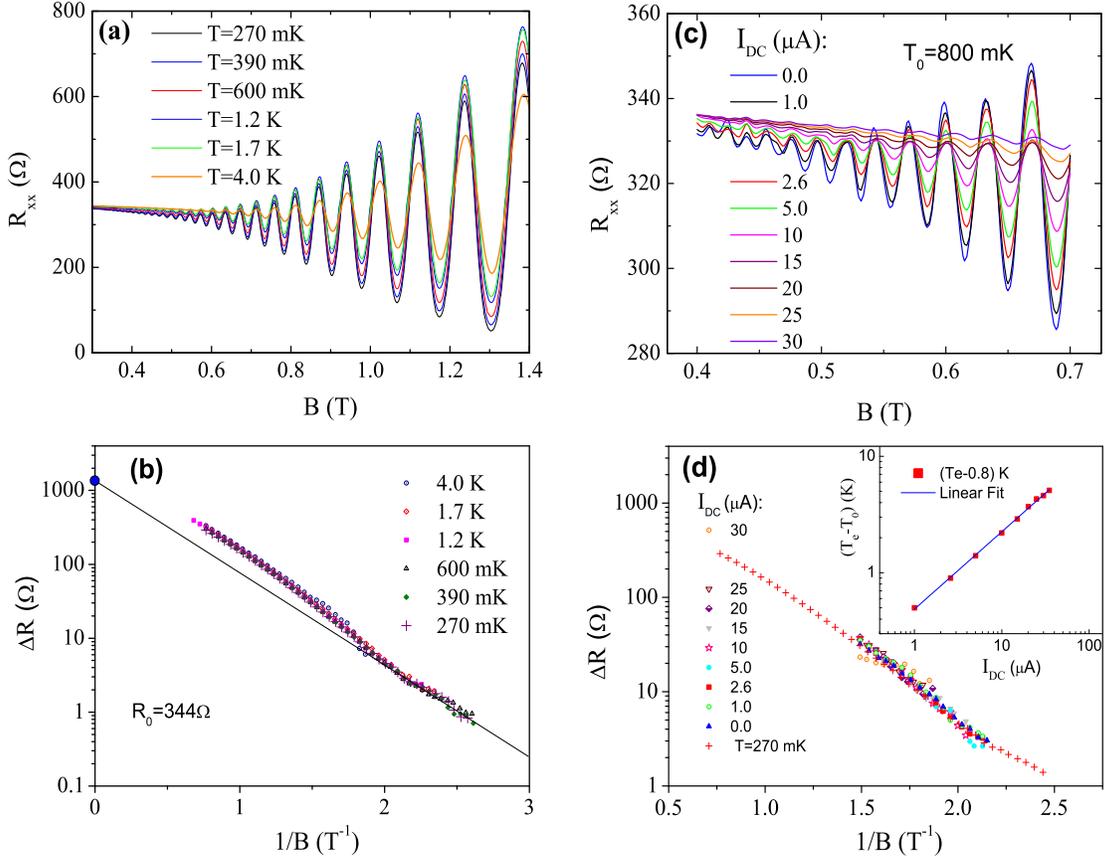}
\caption{(Color online) Calibration measurements of SdH oscillations at different temperatures and dc magnitudes to determine the effective electron temperature $T_e$ as a function of $I_{\texttt{dc}}$.
\textbf{(a)} SdH traces of resistance, $R_{xx}=V_{xx}/I$ at different bath temperatures.
Traces with smaller SdH amplitude correspond to higher temperatures.
\textbf{(b)} Dingle plots of SdH amplitude vs. 1/B  for all traces in (a); the SdH oscillations amplitude is normalized by the thermal damping factor $D_T$; the straight line is a power-law fit through the low-field data.
\textbf{(c)} SdH oscillations of $R_{xx}$ at different $\Idc$ magnitudes. Traces with smaller SdH amplitude correspond to higher values of  $I_{\texttt{dc}}$.
\textbf{(d)} The  Dingle plots corresponding to SdH  oscillations in Fig. (c) normalized by $D_{T_e}$  with single $T_e$ for each $I_{\texttt{dc}}$ (for details see text). The inset shows the deviation of the electron temperature from the equilibrium lattice temperature $T_e-T_0$ as a function of dc current in log-log scales, $T_0$=0.8 K. }
 \label{Fig2abcd}
\end{figure}

As a first step in our analysis it is necessary to determine electron temperature, $T_e$, as a function of $I_{\texttt{dc}}$.
Figure 2~(a) contains SdH traces at different temperatures obtained by sweeping the magnetic field.
Figure 2(b) shows the corresponding Dingle plots\cite{Coleridge} of the SdH amplitude divided by the thermal damping factor $D_T$ vs. inverse magnetic field.  All of the data taken at different temperatures collapse onto a single line after normalization by $D_T$, confirming the validity of eq.~(1), which therefore can be used to extract the electron temperature under the non-equilibrium conditions produced by $I_{\texttt{dc}}$.
The straight line fitted through the low-field data in plot (b) intercepts the ordinate axis at the correct theoretical value of $4R_0$ marked with a solid circle.\cite{Coleridge}
A small deviation of the data from the straight line  seen in Fig.~2 (b) at larger magnetic fields (smaller 1/B) is often observed in experiments and may be attributed to various factors, such as an onset of the spin splitting, spin-orbit effects,  DOS deviation from an ideal shape, and possible small non-uniformities.\cite{Coleridge,coleridge2,studenikinSO}
This small deviation  is not the focus of the current paper as we use the SdH amplitude damping solely for the $T_e$ calibration.

The plots in Fig.~2 (c) show SdH oscillations in resistance (not the derivative) for different $I_{\texttt{dc}}$ values measured at T=800 mK.  A higher bath temperature was chosen to suppress the ZCA effect, which will be discussed separately.
The data in Fig.~2 (c) are extracted from a grey-scale plot similar to the one in Fig. 3 obtained by sweeping $I_{\texttt{dc}}$ and stepping magnetic field, therefore, the resolution vs. B is somewhat smaller than in Fig.~2 (a) but still sufficient for the purpose of  $T_e$ calibration.

\begin{figure}
\includegraphics[width=0.8\columnwidth]{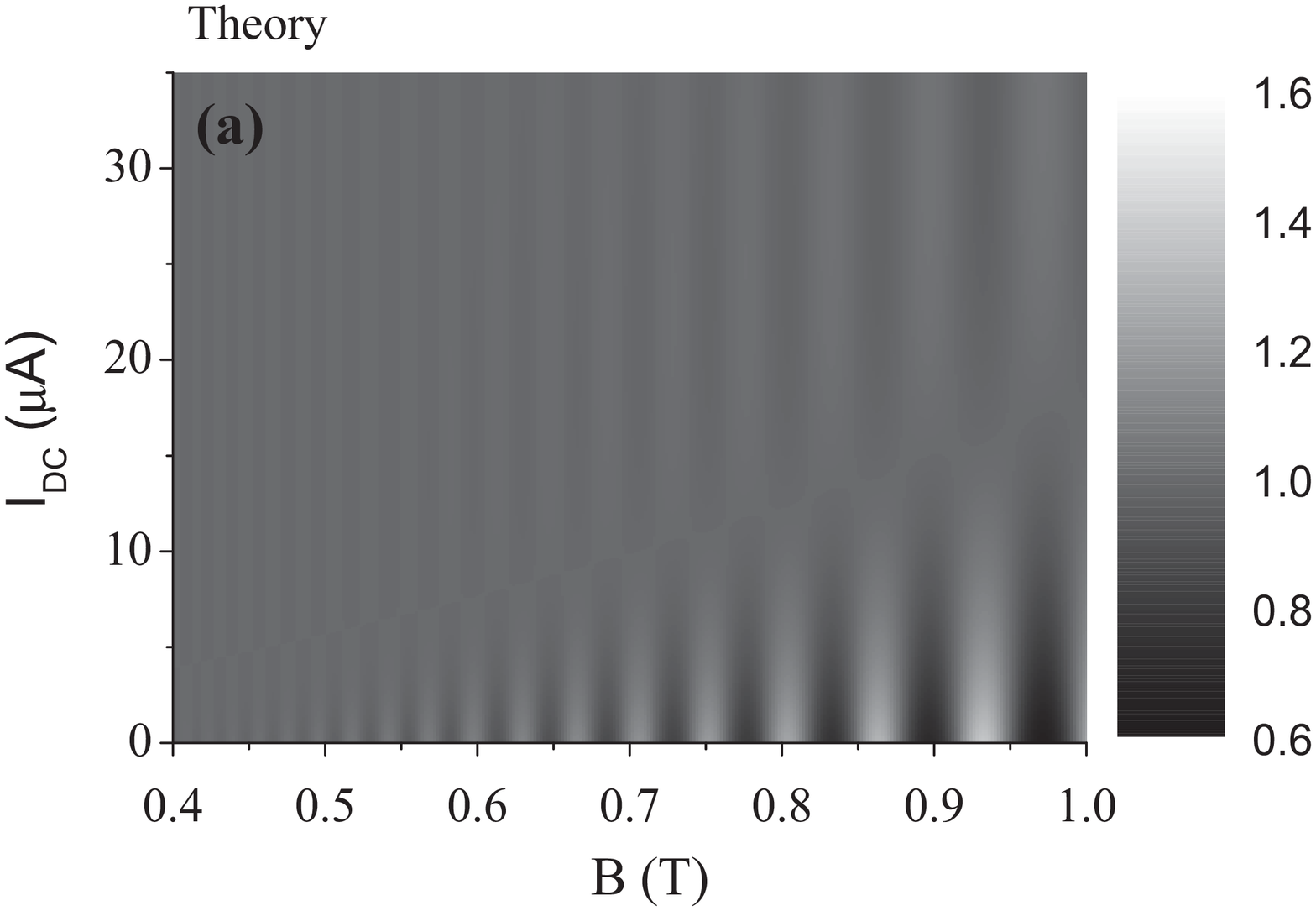}
\includegraphics[width=0.8\columnwidth]{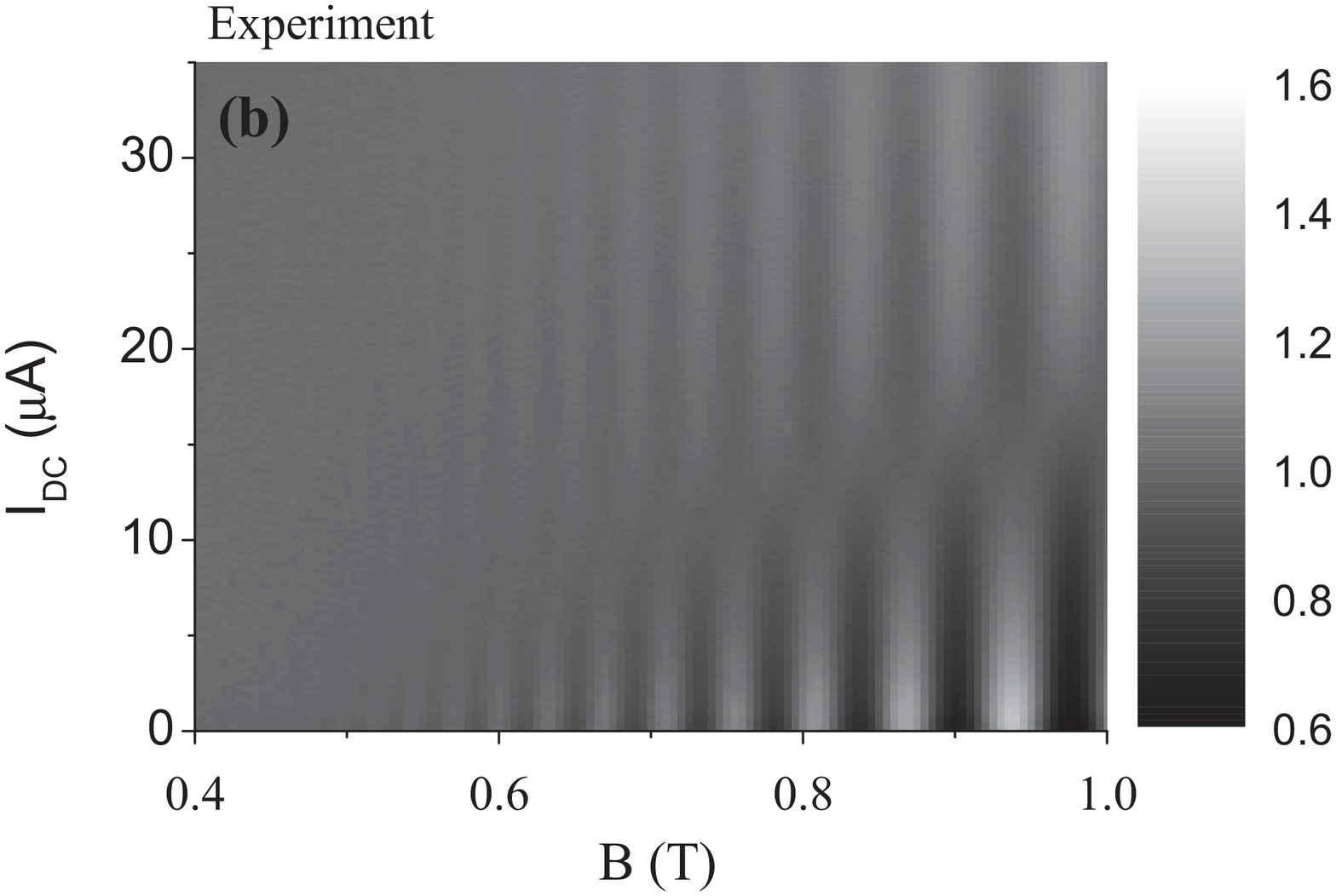}
\caption{Grey-scale plots of the simulated (a) and experimental (b) differential resistance
$r_{xx} = \partial V_x/\partial I$ vs. magnetic field and dc current. Experimental data are taken at 0.8~K by sweeping $I_{\texttt{dc}}$ and stepping magnetic field. Simulation is based on eqs. (1,2) using the empirical calibration $T_e(I_{\texttt{dc}})$ in Fig.~2 (d) and the LL width $\Gamma_L$=1.0 meV obtained from Fig.~2 (b). Color scales are in units of k$\Omega$. }
 \label{Fig3new}
\end{figure}

As expected,  SdH oscillations in Fig. 2~(c) are increasingly damped with increasing dc bias.
The amplitude of SdH oscillations at different $\Idc$ vs. inverse magnetic field is plotted in the main panel of Fig.~2 (d). The curve labeled as T=270 mK (crosses) is taken from Fig.~2 (b) to confirm that the two data sets coincide.
 All the data points fall onto a single line  in Fig. 2 (d) after being divided by  $D_{T_e}$ with the only adjustable parameter  $T_e$.
The values of $T_e$ determined this way  are plotted in the inset of Fig. 2(d) as solid squares in log-log scales. The fitted solid line in this graph is given by the following power-law equation  $\Delta T_e=0.487 I_{\texttt{dc}}^{\alpha}$, where $\Delta T_e = T_e-T_0$ with $T_0$ being the bath temperature in Kelvins, $I_{\texttt{dc}}$ in $\mu A$, and the dimensionless exponent  $\alpha=2/3\pm0.01$.   Noting that the Joule power dissipated in a resistive sample is proportional to $R_0 I_{\texttt{dc}}^2$,  the above relation  $\Delta T_e \propto I_{\texttt{dc}}^{2/3}$  can be reformulated such that the power dissipated as heat by the 2DEG in our experiment is proportional to the third power of electron temperature, $P \propto (T_e-T_0)^3.$   This result is consistent with previous works on the electron energy-loss rates measured in similar 2DEG samples.\cite{fletcher2,leadley}  For example, Fletcher et al.\cite{fletcher2}  found that the experimental energy-loss rate is proportional to the cubic power of the electron temperature for lattice temperatures above 1K.
It should be mentioned that a linear dependence of electron temperature vs. $I_{\texttt{dc}}$ was reported in the much lower mobility graphene sample.\cite{Tan11}

Using the calibration  dependence $T_e(I_{\texttt{dc}})$ we can now numerically simulate the differential resistance in the $I_{\texttt{dc}} - B$ plane using eq. (1) and the  following relation:\cite{Tan11}
\begin{equation}
r_{xx} \equiv \left (\frac{\partial V_x}{\partial I}\right )_{I_{\texttt{dc}}}=  R_{xx}+ I_{\texttt{dc}} \frac{\partial R_{xx}}{\partial T_e}\frac{\partial T_e}{\partial I}    ,
\label{eq.2}
\end{equation}
where $V_x=IR_{xx}$ is the voltage drop measured between potential probes, $R_{xx}=R_0 +\Delta R_{xx}$ is the sample resistance in normal magnetic field defined by eq.~(1) which depends on the electron temperature through the thermal factor $D_T$, and $T_e = T_0 + 0.487 I_{\texttt{dc}}^{2/3}$ is the empirical dependence obtained above with $T_e$ in K and $I_{\texttt{dc}}$ in $\mu A$. The Dingle parameter determined from Fig.~2 (b) is $D_q$=6.3~meV, which is equivalent to a Lorentzian LL width $\Gamma_L=D_q/2\pi$=1.0~meV, or to a quantum scattering time $\tau_q$ of 0.23 ps ($\Gamma_L = \hbar/2\tau_q$).

Figure 3 shows the simulation results from the model (a) and the experimental data (b) of the differential resistance as a grey-scale plot vs. magnetic field and dc current.
Excellent agreement is evident  between theory and the experiment confirming that electron gas heating effect can well account for the observed SdH phase inversion.
Qualitatively, the SdH phase inversion stems from the second term in eq.~(2), which can attain negative sign  due to the  negative derivative $\partial R_{xx}/\partial T_e$ at the SdH maxima.
  In the next section we will extend our study  to higher magnetic fields, in the QHE regime.

\section{Measurements in quantizing magnetic fields.}
Figure 4~(a) presents $r_{xx}$ measurements as a grey-scale plot vs.~normal magnetic field and dc current similar to Fig.~3 but in quantizing magnetic fields when LL levels are well separated, spin minima are resolved, and $R_{xx}$ reaches zero at the SdH oscillations minima.
Comparing Figs. 3 and 4 it is evident that the $r_{xx}$ behavior has qualitatively changed  from a stripe-like pattern to diamond-shaped structures in QHE regime with linear-slope boundaries developing from the SdH minima.
It is evident that the odd spin-gap diamonds  close much faster in Fig.~4 in agreement with expected  smaller spin spitting gaps as compared to the cyclotron ones.
It is worth  mentioning at this point that, while diamond-like features have not been previously reported in the grey-scale form like in Fig.~4, nevertheless, some evidences of such a behavior  can be found in earlier experiments on the QHE breakdown.\cite{kuchar,nachtwei,jeckelmann,kawaji}
The QHE breakdown phenomenon develops as a sudden increase of the  magnetoresistance $R_{xx}=V_{xx}/I$ when current increases above certain critical value.  After differentiation, data presented in the above references\cite{kuchar,nachtwei,jeckelmann,kawaji}   would resemble the transport diamonds under discussion.
In our work we extend earlier observations from the QHE regime to small fields and find continuous evolution between SdH phase inversion and the  QHE breakdown phenomena.
Further experimental and also theoretical studies of such an evolution may provide additional insights into the QHE breakdown phenomenon which is still not fully resolved.\cite{kuchar,nachtwei,jeckelmann,kawaji}

\begin{figure}
\includegraphics[width=0.8\columnwidth]{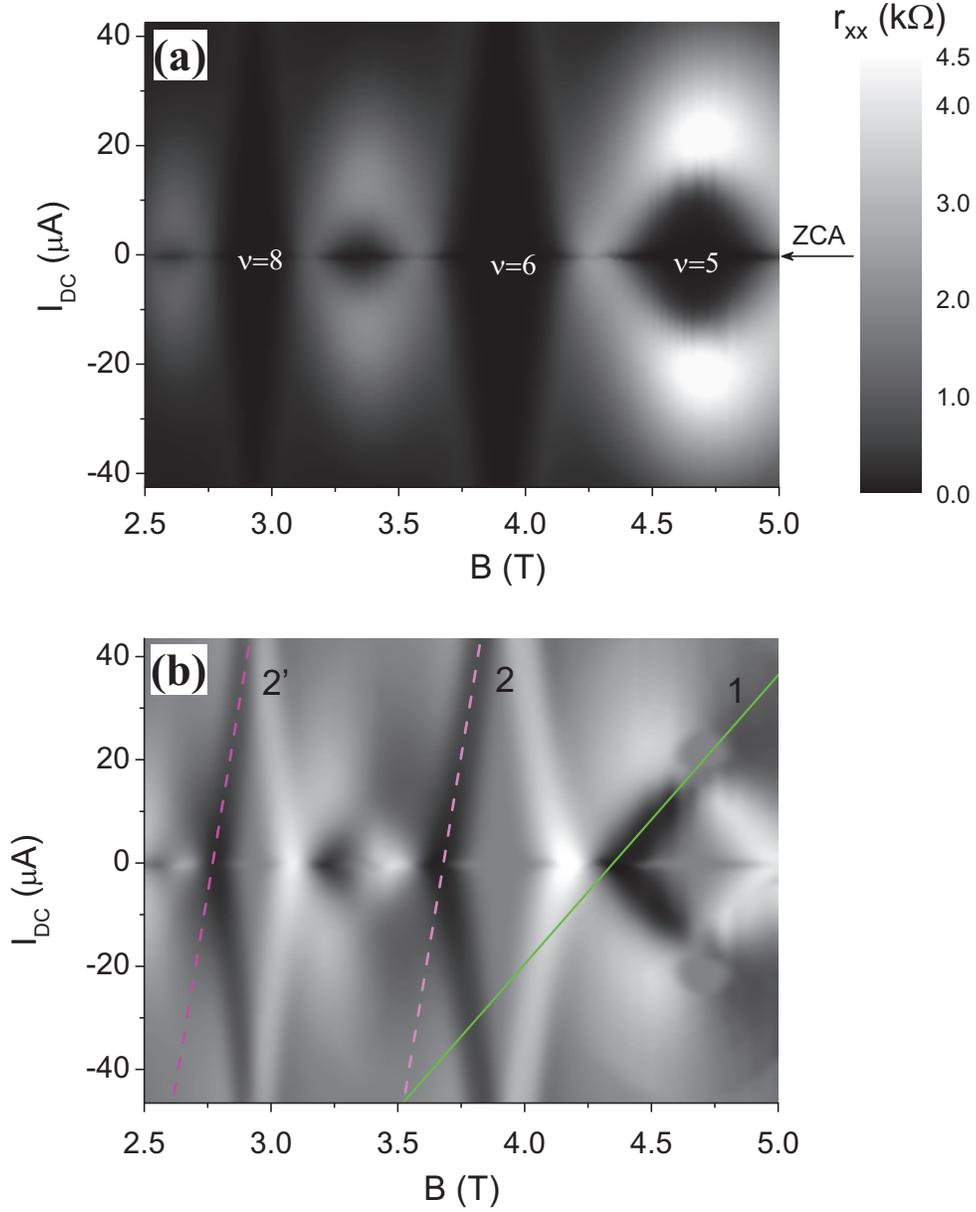}
\caption{\textbf{(a)} Experimental differential resistance $r_{xx}=dV_{xx}/dI$ measured on  an InGaAs/InP Hall bar structure (w=100~$\mu m$) plotted as a grey-scale plot vs. normal magnetic field and dc current. The ZCA position at $\Idc$=0 is indicated by the arrow.
\textbf{(b)} A  derivative with respect to the magnetic field of the data in (a) smoothed over 3 points, $dr_{xx}/dB=d^2V_{xx}/dBdI$.  Color scale is in arbitrary units.  Solid  and dash lines are best fits along the diamond slopes: for $\nu=5$ (solid line 1) $\Idc$=56(B-4.35); for $\nu=6$ (dash line 2) $\Idc$=295(B-3.68), and line 2' has the same slope as line 2 but translated horizontally, $\Idc$=295(B-2.77), where $\Idc$ is in $\mu$A and B is in Tesla.  The slopes  are used to estimate the effective g-factor (for details see text).
 }
 \label{Fig4}
\end{figure}

Based on the above observation that the spin-gap diamonds have much smaller slopes let us estimate the effective electron g-factor adopting an approach developed in  ref. \onlinecite{kawaji}.
 To make diamond edges more visible, Fig.~4(b) presents the derivative with respect to the magnetic field of the smoothed data from Fig.~4(a),  $dr_{xx}/dB=d^2V_{xx}/dBdI$.  Straight lines 1 and 2 in this figure are fitted through the diamond edges at neighboring filling factors $\nu$= 5 and 6, respectively. Line 2' has the same slope as line 2 and is plotted to facilitate the comparison of the slopes at different filling factors, $\nu=$6 and 8, which do not noticeably change with magnetic field.\footnote{In this work we are limited to B$ \leq $ 5T and could not verify the diamond slopes variation over a wider range of quantizing magnetic fields.}

We estimate the effective g-factor from the ratio of the transport diamond slopes assuming that the slope angles are proportional to the corresponding gaps, $\hbar \omega_c$ and $g^*\mu_B B$.
  Such an approach can be justified by making the reasonable assumption that the same mechanism is ultimately responsible for the diamonds at both even and odd filling factors, which differ only by their energy gaps.
 Using the ratio of slopes in Fig.~4(b)  and the electron effective mass  from ref.\cite{m*,martinez} $m^*= 0.047 m_0$, where $m_0$ is the free electron mass, we arrive at the following value for the effective electron g-factor at B=4.7 T ($\nu$=5):  $g^*$=8.2.
 The theoretical value of g-factor for our In$_x$Ga$_{1-x}$As/InP QW sample\cite{ghislain,g*} (x=0.76, w=10~nm, $E_F$=27~meV) is $\mid g(E_F)\mid$=5.45.
 This value is smaller than the experimental $g^*$=8.2 obtained from Fig.~4 and is not surprising because an enhanced g-factor is often observed in magnetotransport measurements due to many-body electron-electron  exchange interactions.\cite{leadley,nicholas}
 The spin-split enhancement is relatively larger in systems with smaller bare values of g-factor, such as GaAs which will be discussed later. In the case of InGaAs QW samples of a similar composition and electron concentration to ours, ref.\cite{desrat} reports similar effective g-factors in a range from 6.5 to 9 obtained from magnetotransport measurements using the LL coincidence method in tilted magnetic fields.

To evaluate the activation energy gap of the $\nu$=5 spin split minimum, we have also performed temperature measurements of $R_{xx}$.  From temperature activation measurements (not shown) we extract g*=1.4, smaller than the g-factor obtained above from the corresponding spin diamond slopes. We attribute this difference to the fact that $\nu$=5 is, actually, the first well resolved spin-split minimum (n.b. higher magnetic fields were not available) at which LL broadening is still comparable with the spin gap.  Specifically,  in temperature activation experiments one measures an effective energy gap between Landau levels, which also involves the  Landau level width, $\Gamma$:  $E_a = g^*\mu_B B - \Gamma $.   From the above results it appears  that the diamond slopes depend only on the B-dependent $\hbar \omega_c$ and $g^*\mu_B B$  values and therefore $\Gamma$ may be excluded. This empirical assumption will require further theoretical confirmation, which is beyond the scope of this article.

\begin{figure}
\includegraphics[width=0.8\columnwidth]{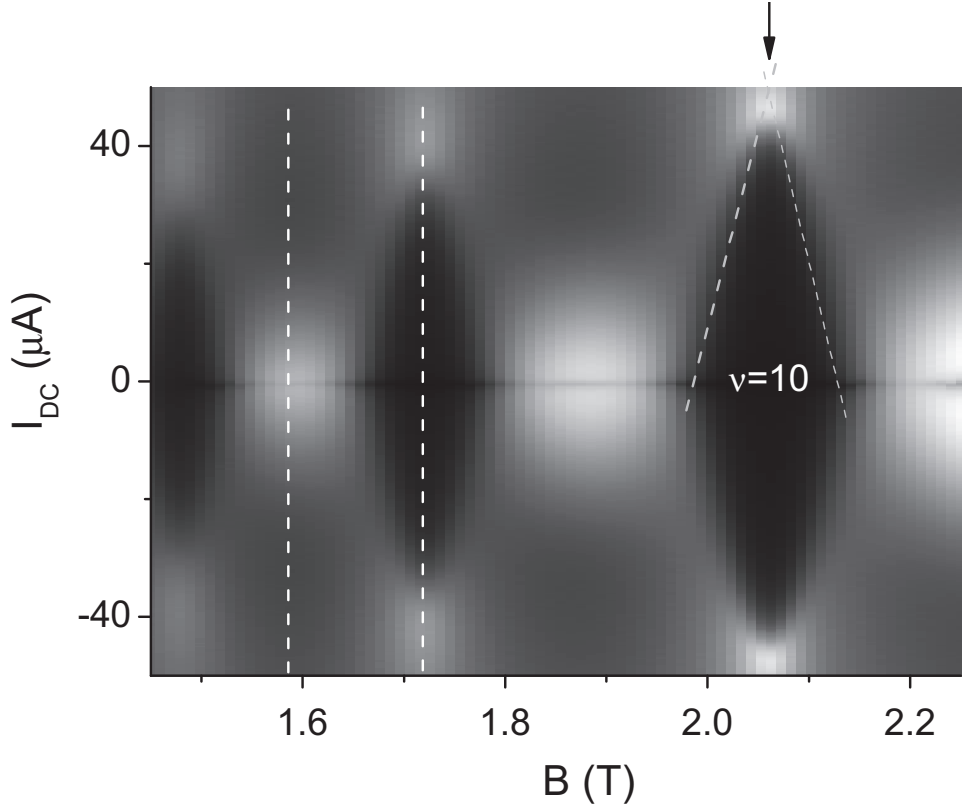}
\caption{ Grey-scale plot of the differential magnetoresistance $r_{xx}$ vs. magnetic field and dc bias in an InGaAs/InP Hall bar structure (w=100$\mu$m) at T=300 mK (black is encoded as zero, white is high).  A sharp maximum in $r_{xx}$ occurs when two opposite diamond edges intersect (marked by the arrow).  The $\nu$=10 diamond is used to estimate the slope angle in units of electron energy per unit dc, $d(eV)/dI \approx 50~\mu eV/\mu A$.
 }
 \label{Fig5}
\end{figure}

\section{Zero Current Anomaly.}
In Fig.~4 there is a peculiar feature near zero dc current that we refer to as a zero current anomaly effect, ZCA, which has has not been previously reported in magnetotransport measurements.
Figure 5 presents another example of the transport diamonds and ZCA in an intermediate  range of magnetic fields, when the QHE regime is not fully reached and the spin sublevels are not yet resolved.
 The ZCA dip is observed over a wide range of quantizing fields independent of the filling factor, i.e.  at SdH maxima  and also at minima if the resistance has not reached zero.   With increasing magnetic field when $r_{xx}$ approaches zero, the ZCA dip first becomes progressively wider (as for example shown in Fig.6~(a) at B=1.72 T) and eventually disappears because $r_{xx}$ simply remains zero in  the strong QHE regime.

 \begin{figure}
\includegraphics[width=0.8\columnwidth]{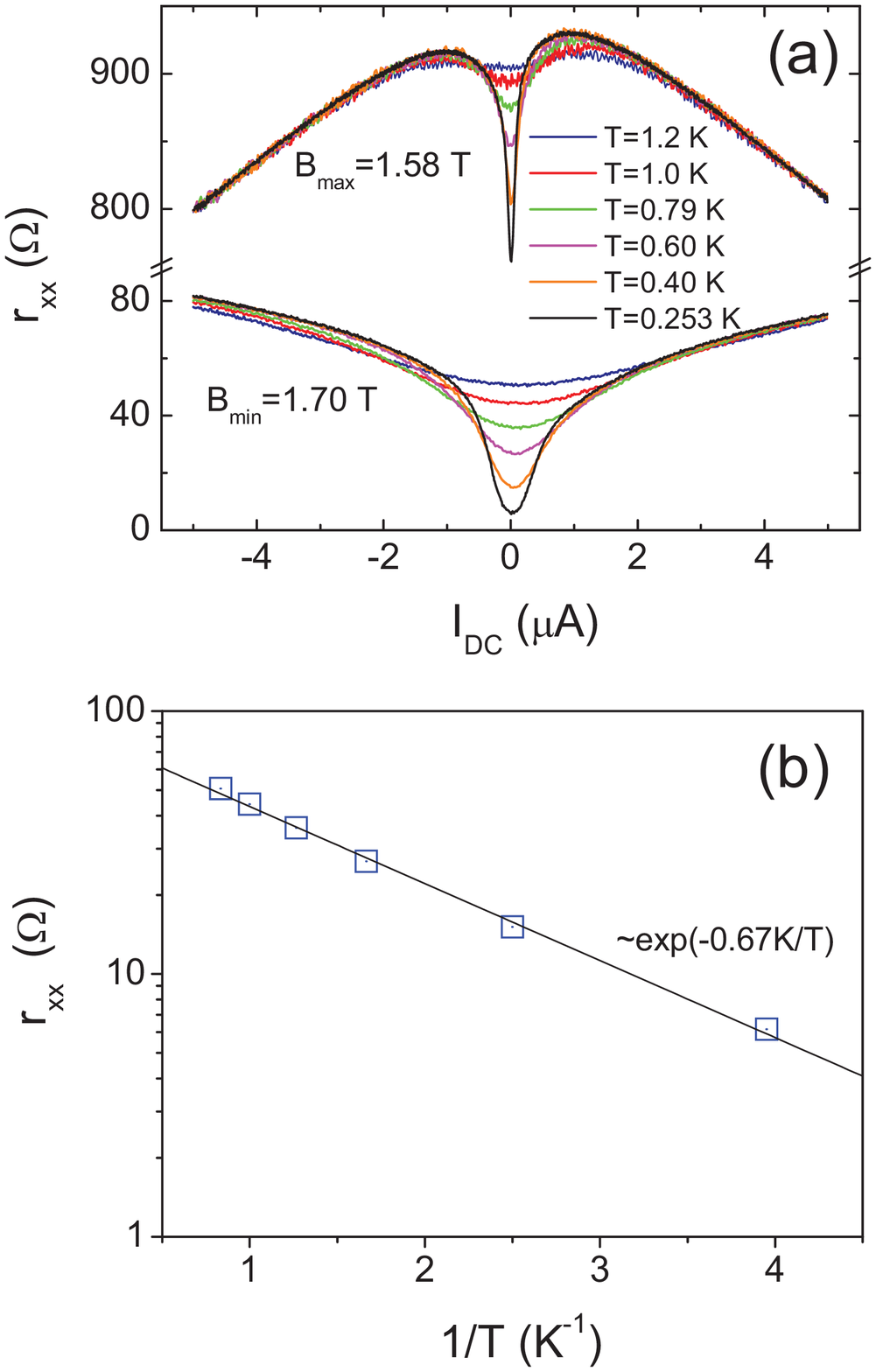}
\caption{\textbf{(a)}  Line traces of the ZCA   effect for different temperatures at two magnetic field values corresponding to a maximum $B_{max}$=1.58 T and a minimum $B_{min}$=1.72 T of SdH oscillations indicated in Fig. 5 by vertical dashed lines.
\textbf{(b)} Arrhenius activation plot of  the $r_{xx}$ data in (a)  at B=1.72 T (squares). The solid line is the best exponential fit with the activation energy of 0.67 K.
 }
 \label{Fig6}
\end{figure}

Examples of the ZCA line traces at different temperatures are shown in Fig.~6~(a) for two magnetic field values corresponding to a minimum and a maximum of the SdH oscillations indicated by vertical dashed lines in Fig.~5.
It is evident from Fig. 6~(a) that the ZCA  strongly depends on temperature.  Qualitatively,  the ZCA has a similar temperature behavior  at all filling factors vanishing at temperatures higher than 1~K.
 As a theoretical description of ZCA is not available, for an estimation of the activation energy we adopt a similar approach used in the  QHE regime: we choose a SdH minimum where the ZCA dip is close to zero at the lowest temperature (B=1.72~T in Fig.~6(a)) and  plot $r_{xx}$ in logarithmic scale as a function of inverse temperature.  This activation plot is presented in Fig.~6(b).  The solid line gives an exponential fit $\propto \exp(-\Delta/T)$ with $\Delta$ = 0.67~K, or 58 $\mu$eV in energy units.  The cyclotron gap at B=1.72T is 4.2 meV, which is almost two orders of magnitude larger than $\Delta$.  
  In the following we compare this value of $\Delta$ to that obtained using an alternative approach.
 
  As is evident from  Figs.~4 and 5, the diamond slopes (once formed) do not noticeably change with  magnetic field and therefore the cyclotron gap diamond slopes can be used to estimate the energy scale of the ZCA effect.
In order to estimate the ZCA width in energy units we first calibrate the effective electric field acting upon conduction electrons.
We use the cyclotron gap diamond $\nu$=10 in Fig.~5 for this purpose.
The linear dependence of the diamond edges  indicates that the effective field  is proportional to the  magnitude of $\Idc$.
The  $\nu$=10 diamond (at B=2.06 T and highlighted in Fig. 5 by dashed lines) closes  at $\Idc\approx$ 50 $\mu$A.   The electron effective mass $m^*$=0.047$m_0$,\cite{martinez} gives us the cyclotron energy  $\hbar \omega_c$= 5.1~meV.  
Using these numbers, from Fig.~5   we calculate the slope angle in units of electron energy per unit current  $edV^*/dI_{\texttt{dc}} \approx$50 eV/A  and  a corresponding effective electric field  $E^* =  V^*/\texttt{w} \approx 0.5 \Idc$, where $E^*$ is in V/m, and $I_{\texttt{dc}}$ in $\mu$A.
Having this slope calibration we can  estimate now the width of the ZCA dip in $\mu$eV.
   For the data presented in Fig.~6(a) at a SdH maximum (B$_{\textrm{max}}$=1.58 T) the FWHM  $\Delta_{ZCA}\approx$ 3 $\mu$A, which according to the above calibration is equivalent to 150 $\mu e$V.  Note that this estimate is valid at maxima of SdH oscillations where the current density and electric field are uniformly distributed across the Hall bar width.
 At SdH minima when $r_{xx}$ is close to zero, the above estimate may not be valid due to edge state effects leading to a non-uniform distribution of electric field. This is consistent with a wider ZCA dip in Fig.~6(a) at the SdH minima where the resistivity is close to zero.

\begin{figure}
\includegraphics[width=0.8\columnwidth]{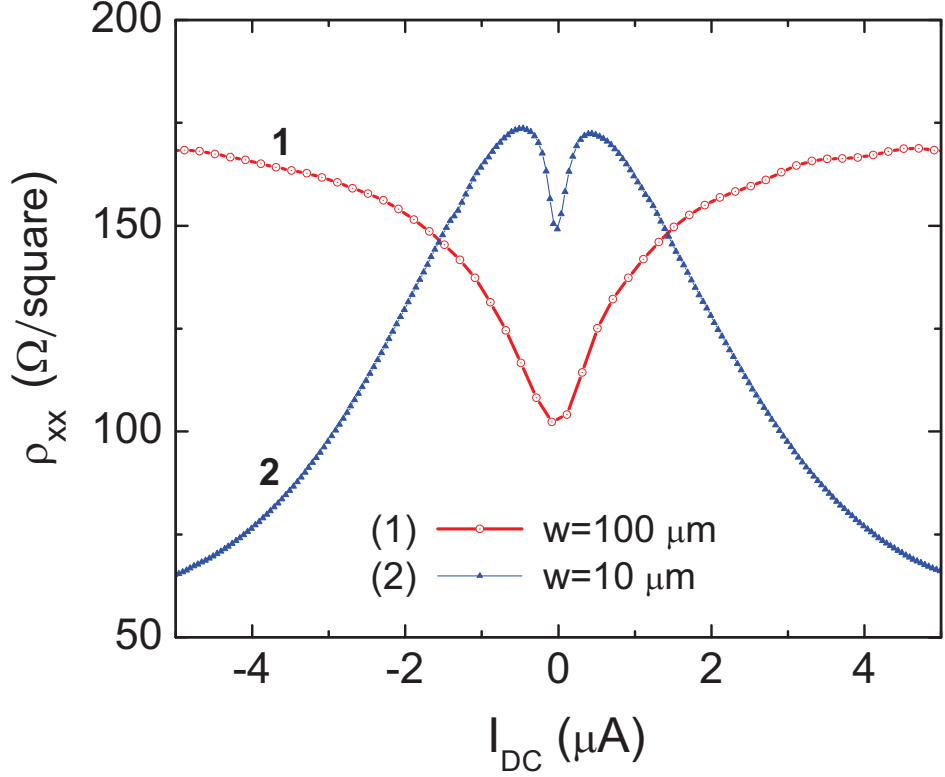}
\caption{Differential resistivity $\rho_{xx}$ in $ \Omega$/square measured in the same experiment on InGaAs/InP stripes of different widths (depicted in Fig. 1 (b)) at T=270~mK and B=2.2 T; (1) w=100~$\mu m$ and (2) w=10~$\mu m$. }
 \label{Fig7}
\end{figure}

  We have verified that the ZCA is not a contact effect by performing 4-point, 3-point and 2-point measurements using different sets of contacts, from which we conclude that the contact resistance is small in comparison with the Hall bar resistance.   All measurements reveal a narrow ZCA.
  The ZCA effect is observed in several experiments on Hall bars of different widths from 200 to 10~$\mu m$ made from different materials, including InGaAs/InP, InAsP/InP (not shown) and GaAs/AlGaAs (presented below).
  Figure 7 demonstrates the ZCA effect on InGaAs/InP strips of different widths, 100 and 10~$\mu$m, measured in the same experiment.  The 100~$\mu$m strip corresponds to the main Hall bar labeled (1) in Fig. 1, while a narrow potential arm of 10~$\mu$m width is used in the second measurement (2). It is evident from Fig.~7 that the ZCA width is inverse-proportional to the Hall bar width meaning it is driven by dc density or by the Hall field. Further studies are  necessary to distinguish between the two possibilities.

 \section{Nonlinear transport measurements in a $\mbox{GaAs/AlGaAs}$ heterostructure.}
\begin{figure}
\includegraphics[width=0.8\columnwidth]{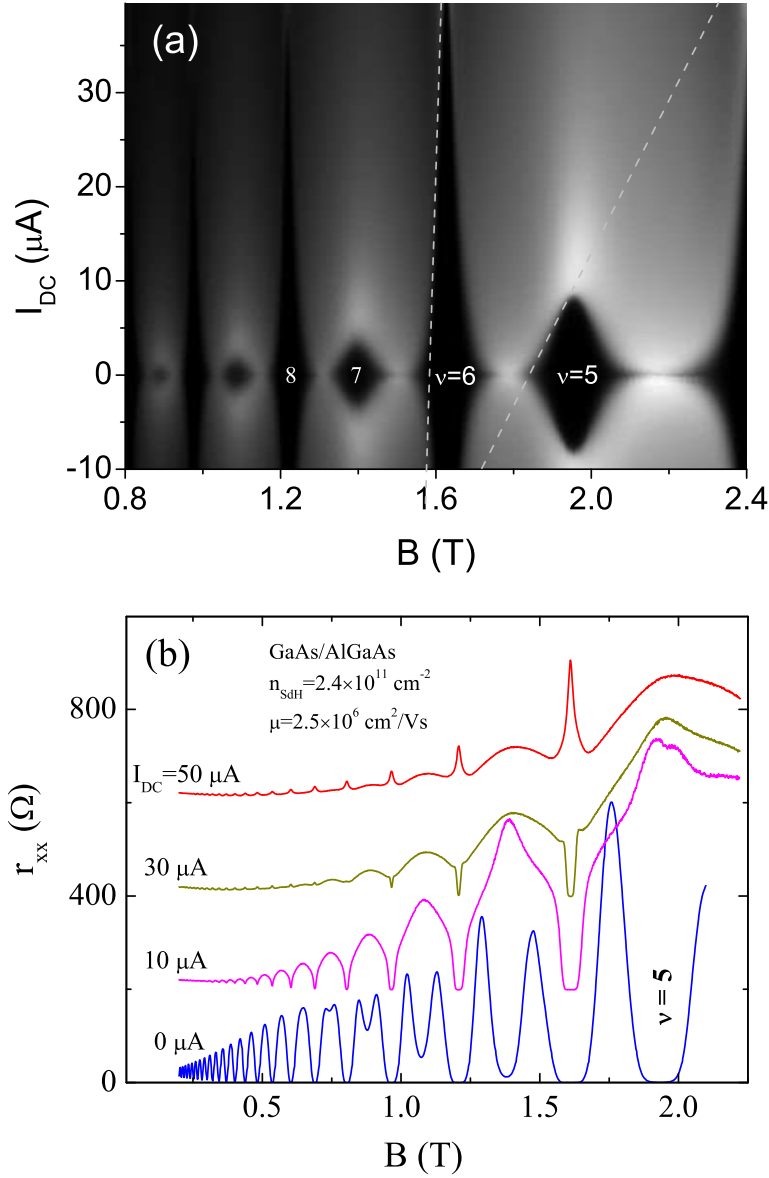}
\caption{Differential resistance  measurements on a GaAs/AlGaAs Hall bar sample (w=200~$\mu m$) at T=270~mK: (a) grey-scale plot of $r_{xx}$ vs. magnetic field and dc current; zero is encoded as black.
(b) Individual magnetic field traces at different values of $\Idc$. }
 \label{Fig8}
\end{figure}
In order to find out whether the observed phenomena are related to specific properties of the InGaAs/InP material system, e.g. related to its large electron effective g-factor and spin-orbit coupling, we have conducted a similar experiment on a Hall bar (w=200~$\mu m$) fabricated from a standard high mobility GaAs/AlGaAs heterostructure.   The 2DEG in this structure had a density of 2.4$\times 10^{11}$ cm$^{-2}$ and a mobility of 2.5$\times 10^6$ cm$^2$/Vs.
The results of these measurements are presented in Fig.~8 as a grey-scale plot in (a),  and  as individual SdH traces at different dc values in (b).
It is evident from Fig. 8 that the features seen in InGaAs  are also present in nonlinear magnetotransport measurements on the GaAs/AlGaAs structure, namely, transport diamonds at even/odd filling factors, a sharp anomaly (ZCA) at zero dc, and the phase inversion of SdH oscillations in lower magnetic field range.  The observed  features are sharper in the GaAs/AlGaAs structure due to much higher electron mobility.

 From the ratio of the diamond slopes marked in Fig.~8(a) by dashed lines, and  knowing the electron effective mass in GaAs $m^*=0.067 m_0 $, we estimate g$^*$=1.9 at $\nu$=5.
 Again, this value is larger than the bare $|g|$=0.44 g-factor in GaAs due to electron-electron exchange interactions as discussed above.\cite{leadley,nicholas}
  From the temperature activation measurements (not shown) of the corresponding SdH minima ($\nu$=5) we  obtain g*=1.8.  The two values of g-factor obtained from transport diamonds and from the activation plot are now much closer to each other than those measured in InGaAs/InP.
  This is related to the much higher electron mobility in the GaAs/AlGaAs sample and, consequently, a smaller Landau level width, $\Gamma$,  leading to a smaller correction to the activation energy gap $E_a=g^*\mu B - \Gamma$.
 From  $\nu$=6 diamond in Fig.~8(a) and  the value of the cyclotron energy at this field  we calibrate the slope  $edV^*/dI_{dc}$=62~eV/A.  The obtained  value of $edV^*/dI_{dc}$ in GaAs/AlGaAs sample   is very close to the one obtained in InGaAs, even though mobilities in these samples differ by more than a factor of 10. Note that the electron concentration in the studied  GaAs/AlGaAs sample is roughly twice smaller than in the InGaAs/InP one.  This difference is compensated by a twice wider GaAs Hall  bar to produce roughly the same Hall field for the same dc value. This important observation indicates that the transport diamonds are driven by the non-dissipative component of the conductivity tensor, which does not depend on the transport relaxation time.

It is evident from Fig.~8(a) that the ZCA effect is also present in the GaAs /AlGaAs sample, although it is somewhat narrower than in InGaAs, apparently, due to smaller disorder in a higher mobility sample leading to a smaller ZCA activation energy.  Indeed, applying the same procedure as in the case of InGaAs/InP above, we find $\Delta_{\textrm{ZCA}}\approx$ 40 $\mu e$V at B=1.51 T, which is approximately 4 times smaller than the value of $\Delta_{\textrm{ZCA}}$  obtained in InGaAs.
This indicates that the ZCA width is sensitive to disorder (potential fluctuations). The ZCA in a higher mobility GaAs/AlGaAs structure is expected to become more pronounced at lower temperatures when the thermal energy becomes smaller than  the corresponding activation energy $\Delta_{\textrm{ZCA}}$.

Since the ZCA effect is observed in both InGaAs and GaAs samples it may be  concluded that the ZCA is a fundamental property of a 2DEG.
It can possibly  originate from a Coulomb gap in the one-particle DOS of interacting electrons in the presence of disorder and magnetic field.\cite{fogler,burmistrov}
Such a DOS anomaly has been observed as a zero-bias anomaly (ZBA) in 2D tunneling experiments. \cite{ashoori,dial,reker,khanin}    Another phenomenon that has been recently actively discussed in the literature is the ZCA in the I-V characteristics of quantum point contacts (QPC).\cite{cronenwett,pepper,frolov}  The ZBA in QPC experiments has very similar characteristics compared to the ones reported in this work, e.g. in ref.~\onlinecite{frolov} the ZBA has a width of $\approx$100~$\mu e$V and has a similar temperature dependence vanishing above 1~K.
 Therefore, the results reported in this work may provide additional insights into the origin of these separate, but perhaps related, phenomena.

\section{Conclusions}
In summary,  we have studied nonlinear transport  phenomena in  Hall-bar samples prepared from InGaAs/InP and GaAs/AlGaAs  structures containing a high-mobility 2DEG.
Two non-linear phenomena, SdH phase inversion and ZCA, have been observed in both the material systems.
It is shown that the SdH phase inversion can be well described by Isihara's equation\cite{Isihara} taking into account electron heating effects.\cite{Tan11}     A stripe-like pattern of the SdH inversion effect evolves into diamond-shaped structures in the QHE regime with linear slopes.
The observed transport diamonds at even and odd filling factors  can be used for gaining information about the effective electron g-factor and, in principle, for probing other many-particle gaps; e.g. we speculate that such a technique may be used for testing energy gaps in the fractional quantum Hall regime.
The ZCA effect is another curious phenomenon to be understood.  This phenomenon is observed as a sharp dip in $\rxx$ vs. $\Idc$ at temperatures below 1~K in a wide range of quantizing magnetic fields.
It will be interesting to find out whether the observed ZCA phenomenon is linked to other 2DEG experiments where a zero bias anomaly is observed.\cite{ashoori,dial,reker,cronenwett,frolov,majorana}

\section{Acknowledgements}
 We thank S. Vitkalov, P. Coleridge, Ch. Dharma-Wardana, and O. Dial for their interest and helpful discussions. G.G. acknowledges financial support from the CNRS-NRC collaboration.

\end{document}